\documentclass[a4paper,11pt]{article}
\usepackage{jheppub} 
\usepackage{lineno}
\usepackage{booktabs}
\usepackage{cancel}
\usepackage[title]{appendix}
\usepackage[normalem]{ulem}
\usepackage{orcidlink}

\newcommand{\mevcc}{\,{\ensuremath{\mathrm{\mbox{MeV}}/c^2}}}

\newcommand{\mev}{\,{\ensuremath{\mathrm{\mbox{MeV}}}}}
\newcommand{\gevcc}{\,{\ensuremath{\mathrm{\mbox{GeV}}/c^2}}}
\newcommand{\gevc}{\,{\ensuremath{\mathrm{\mbox{GeV}}/c}}}

\newcommand{\mbs}{M(B_s)}

\newcommand{\mds}{M(D_s)}
\newcommand{\mdp}{M(D^+)}
\newcommand{\mdz}{M(D^0)}

\newcommand{\tabitem}{~~\llap{\;}~~}

\newcommand{\Br}{\ensuremath{\mathcal{B}}}

\title{\boldmath Measurement of the inclusive branching fractions for $B_s^0$ decays into $D$ mesons via hadronic tagging}

\collaboration{The Belle and Belle II Collaborations}
  \author{I.~Adachi\,\orcidlink{0000-0003-2287-0173},} 
  \author{L.~Aggarwal\,\orcidlink{0000-0002-0909-7537},} 
  \author{H.~Ahmed\,\orcidlink{0000-0003-3976-7498},} 
  \author{H.~Aihara\,\orcidlink{0000-0002-1907-5964},} 
  \author{N.~Akopov\,\orcidlink{0000-0002-4425-2096},} 
  \author{A.~Aloisio\,\orcidlink{0000-0002-3883-6693},} 
  \author{S.~Al~Said\,\orcidlink{0000-0002-4895-3869},} 
  \author{N.~Althubiti\,\orcidlink{0000-0003-1513-0409},} 
  \author{N.~Anh~Ky\,\orcidlink{0000-0003-0471-197X},} 
  \author{D.~M.~Asner\,\orcidlink{0000-0002-1586-5790},} 
% ...

\abstract{We report measurements of the absolute branching fractions $\Br(B_s^0 \to D_s^{\pm} X)$, $\Br(B_s^0 \to D^0/\bar{D}^0 X)$, and $\Br(B_s^0 \to D^{\pm} X)$, where the latter is measured for the first time. The results are based on a 121.4\,fb$^{-1}$ data sample collected at the $\Upsilon(10860)$ resonance by the Belle detector at the KEKB asymmetric-energy $e^+ e^-$ collider. We reconstruct one $B_s^0$ meson in $e^+e^- \to \Upsilon(10860) \to B_s^{*} \bar{B}_s^{*}$ events and measure yields of $D_s^+$, $D^0$, and $D^+$ mesons in the rest of the event. We obtain $\Br(B_s^0 \to D_s^{\pm} X) = (68.6 \pm 7.2 \pm 4.0)\%$, $\Br(B_s^0 \to D^0/\bar{D}^0 X) = (21.5 \pm 6.1 \pm 1.8)\%$, and $\Br(B_s^0 \to D^{\pm} X) = (12.6 \pm 4.6 \pm 1.3)\%$, where the first uncertainty is statistical and the second is systematic. Averaging with previous Belle measurements gives $\Br(B_s^0 \to D_s^{\pm} X) = (63.4 \pm 4.5 \pm 2.2)\%$ and $\Br(B_s^0 \to D^0/\bar{D}^0 X) = (23.9 \pm 4.1 \pm 1.8)\%$. For the $B_s^0$ production fraction at the $\Upsilon(10860)$, we find $f_s = (21.4^{+1.5}_{-1.7})\%$.
}

\begin{document}
\maketitle
\flushbottom
\section{Introduction}

Decays of $B$ mesons provide a powerful tool for studying strong interactions at low energy, measuring parameters of the Standard Model, and searching for New Physics~\cite{Kou_2019}. Absolute branching fractions of $B^+$ and $B^0$ decays have been precisely measured by the Belle, BaBar, and Belle~II experiments using $e^+e^-$ collisions at the $\Upsilon(4S)$ resonance. To study $B_s^0$ mesons, Belle collected data at the $\Upsilon(10860)$ resonance, which decays to $B_s^{(*)}\bar{B}_s^{(*)}$, $B^{(*)}\bar{B}^{(*)}(\pi)$, and final states with bottomonium and light hadrons. For brevity, in the following we refer to the $\Upsilon(10860)$ as the $\Upsilon(5S)$. The accuracy of absolute branching fractions of $B_s^0$ decays is limited by our knowledge of the $B_s^0$ production fraction at the $\Upsilon(5S)$ energy, $f_s$. To measure $f_s$, Belle used inclusive production of $D_s^+$ and $D^0$ mesons, to obtain $f_s$ = $(22.0^{+2.0}_{-2.1})\%$~\cite{Belle:2023yfw}. The uncertainty in $f_s$ is dominated by that of the inclusive branching fraction, $\Br(B_s^0 \to D_s^{\pm} X)$, which was recently measured by Belle using semileptonic tagging to be $\Br(B_s^0 \to D_s^{\pm} X) = (60.2\pm5.8\pm2.3)\%$~\cite{Belle:2021qxu}. It is important to measure this branching fraction using hadronic tagging to improve its uncertainty. The sum of $\Br(B_s^0 \to D_s^{\pm} X)$, $\Br(B_s^0 \to D^0/\bar{D}^0 X)$, and $\Br(B_s^0 \to D^{\pm} X)$ is expected to be above 100\%, as the charm quark is produced both in the $b \to c$ and $W^- \to \bar{c} s$ parts of the $B_s^0$ decay diagram, and can be estimated based on similar sums for $B^+$ and $B^0$ mesons (see section~\ref{sumBF}). Thus, measurement of all three branching fractions will allow a consistency check of the results. Recently, Belle measured the ratio, $\Br(B_s^0 \to D^0/\bar{D}^0 X)/\Br(B_s^0 \to D_s^{\pm} X) = 0.416 \pm 0.018 \pm 0.092$~\cite{Belle:2023yfw}, from which we estimate $\Br(B_s^0 \to D^0 /\bar{D}^0 X) = (25.0 \pm 2.6 \pm 5.6)\%$. There is no information on $\Br(B_s^0 \to D^{\pm} X)$.

In this paper, we report measurements of $\Br(B_s^0 \rightarrow D_s^{\pm} X)$, $\Br(B_s^0 \to D^0/\bar{D}^0X)$, and $\Br(B_s^0 \rightarrow D^{\pm}X)$. We use a data sample collected by the Belle experiment at the center-of-mass energy of the $\Upsilon(5S)$ resonance, 10.866~GeV, which has an integrated luminosity of 121.4\,fb$^{-1}$ corresponding to $N_{b \bar{b}}$ = $(41.3 \pm 1.9) \times 10^6$. At the $\Upsilon(5S)$ resonance, $B_s^0$ mesons are produced in the processes $e^+e^- \to B_s^0\bar{B}_s^0$, $B_s^0\bar{B}_s^{*}$, and $B_s^{*} \bar{B}_s^{*}$, with $B_s^* \to B_s^0 \gamma$. We fully reconstruct one $B_s^0$ meson in many hadronic final states using a multivariate full event interpretation (FEI) algorithm~\cite{Keck:2018lcd}. We then reconstruct a $D_s^+$, $D^0$ or $D^+$ meson in the rest of the event (ROE). The branching fraction is calculated as
\begin{equation}
{\cal B}(B_s^0 \to D/\bar{D} X) = \frac{{N_{B_s-D}}}{N_{B_s} \, {\cal B}_{D} \, \varepsilon_{D}^\mathrm{ROE}},
\label{Eq1}
\end{equation}
where $D$ denotes $D_s^+$, $D^0$, and $D^+$, ${\cal B}_{D}$ is the branching fraction of the $D$ reconstruction channel, and $\varepsilon_{D}^\mathrm{ROE}$ is the reconstruction efficiency. The total number of $B_s^0$ tags, $N_{B_s}$, is determined from a fit to the $\mbs$ distribution. The number of $B_s^0 - D$ pairs, $N_{B_s-D}$, is determined from a two-dimensional fit to the distribution in $\mbs$ and $M(D)$, where $\mbs$ and $M(D)$ are the invariant masses of the $B_s^0$ and $D$ candidates, respectively. To avoid bias, the data in the signal region were not examined until the selection criteria were fixed.

\section{Belle detector}
This analysis is based on data collected by the Belle detector
at the KEKB asymmetric-energy $e^+ e^-$ collider~\cite{KUROKAWA20031, KEKB_ach}.
The Belle detector is a large-solid-angle magnetic spectrometer which consists of a four-layer silicon vertex detector (SVD), a 50-layer central drift chamber (CDC), an array of aerogel threshold Cherenkov counters (ACC), time-of-flight scintillation counters (TOF), and an electromagnetic calorimeter (ECL) composed of CsI(Tl) crystals located inside a superconducting solenoid coil that provides a magnetic field of 1.5~T. The $K_L^0$ meson and muon detector (KLM), composed of resistive plate chambers, is located in the iron solenoid yoke. A detailed description of the detector can be found in refs.~\cite{ABASHIAN2002117, Belle_ach}.

Monte-Carlo (MC) simulation of $e^+ e^- \to b \bar{b}$ and continuum $e^+ e^- \to q \bar{q}$ ($q = u, d, s, c$) events uses EvtGen~\cite{LANGE2001152}. The $e^+ e^- \to b \bar{b}$ events are generated from $\Upsilon(5S)$ decays, including $B_s^{(*)}\bar{B}_s^{(*)}$, $B^{(*)}\bar{B}^{(*)}(\pi)$, and final states with bottomonia. The MC sample size corresponds to an integrated luminosity six times larger than the data. The detector response is modeled using GEANT3~\cite{Brun:1987ma}. The MC simulation includes run-dependent variations in detector performance and background conditions.

\section{Event selection}
\subsection{Reconstruction of $B_s^0$ tag candidates}
\label{Bs_sel}

%
We reconstruct $B_s^0$ mesons in the decay channels
$D_s^{(*)-}\pi^+(\pi^0,\,\pi^+\pi^-)$, $D_s^- K^+$, $D_s^{(*)+}D_s^{(*)-}$, $\bar{D}^{(*)0} K^- \pi^+$, and $J/\psi\,K^+ K^- (\pi^0)$.\footnote{Throughout this paper, charge-conjugate channels are
always included.}
The $D^0$, $D^+$, and $D_s^+$ mesons are reconstructed in final
states with $K^\pm$, $K_S^0$, $\pi^\pm$, $\eta$, $\eta'$, up to one $\pi^0$, and up to five decay products.
A list of the channels used in the reconstruction of $B$ and $D$
mesons is given in Appendix~\ref{appendixA}.
We reconstruct $K_S^0$ mesons in the $\pi^+ \pi^-$ channel, $\pi^0$ mesons in the $\gamma \gamma$ channel, $\eta$ mesons in the $\gamma \gamma$ and $\pi^+ \pi^- \pi^0$ channels, $\eta'$ mesons in the $\pi^+ \pi^- \eta$ and $\pi^+ \pi^- \gamma$ channels, $D_s^{*+}$ mesons in the $D_s^+ \gamma$ channel, $D^{*0}$ mesons in the $D^0 \pi^0$ and $D^0\gamma$ channels, $D^{*+}$ mesons in the $D^+ \pi^0$ and $D^0 \pi^+$ channels, and the $J/\psi$ in the $\mu^+ \mu^-$ and $e^+ e^-$ channels. 

We perform an initial loose selection of the final-state particles and decays, and then use a multivariate analysis for the final selection.
We select tracks that originate from the vicinity of the interaction point (IP) by requiring $dr<0.5\,\mathrm{cm}$ and $dz<3\,\mathrm{cm}$, where the $z$-axis is in the direction opposite to the $e^+$ beam, and $dr$ and $dz$ are transverse and longitudinal distances between the track and the IP, respectively.
%
%
Charged particles are identified using ionization energy-loss measurements in the CDC, time-of-flight information from the TOF, and Cherenkov light yields in the ACC. Information from these
subdetectors is combined into a likelihood $L(h)$ for a given hadron hypothesis $h$. In the initial selection, we apply the identification requirement only for kaon candidates, $L(K)$/$(L(K) + L(\pi))$ $>$ 0.1. The efficiency of this requirement is 98\% and the probability to misidentify a pion as a kaon is about 20\%.
%
%
For photons, we require the energy to be greater than $100$, $50$, and
$150\,\mev$ in the forward endcap ($12.4^\circ<\theta<32.2^\circ$),
barrel ($32.2^\circ<\theta<128.7^\circ$), and backward endcap
($\theta> 128.7^\circ$) regions of the ECL, respectively,
as these regions have different levels of background.
For the $\pi^0$, $K_S^0$, $\eta$, $\eta'$, $D$, $D^*$, and $J/\psi$ candidates, we apply a mass range requirement that corresponds to about $\pm 5$ units of mass
resolution. To improve momentum resolution, we apply a mass-constrained fit to $\pi^0$, $\eta$, $D^*$, and $J/\psi$ candidates; a mass-vertex-constrained fit to $\eta'$ and $D$ candidates; and a vertex-constrained fit to $K_S^0$ candidates.

In the FEI algorithm, a boosted decision tree (BDT)~\cite{Keck:2017gsv} is used
with the following discriminating variables for various particle
species:
\begin{itemize}
\item For charged pions, kaons, and leptons, we use the momentum,
 transverse momentum, and particle identification information.
\item For photons, we use the energy, polar angle, number of crystals
 in the energy deposition (cluster), the ratio of the energy
 deposition in a $3\times3$ matrix of crystals to that in a
 $5\times5$ matrix, and cluster timing. These variables suppress hadronic showers and beam background.
\item
For $K_S^0\to\pi^+ \pi^-$ candidates, we use the invariant mass of the $K_S^0$ candidate and a set of parameters describing the displaced vertex of the $K_S^0$. These are the distance of closest
approach between the two daughter pions, the impact parameters of the daughter pions,
the distance between the IP and the $K_S^0$ vertex, and the angle between the $K_S^0$ momentum
and the direction from the IP to the $K_S^0$ vertex; the latter three variables are measured in
the plane perpendicular to the beam direction.
\item
 For $\pi^0\to\gamma\gamma$ candidates, we use the two-photon invariant mass,\footnote{Here and below the invariant mass denotes the mass before the mass-constrained fit.} momentum and decay angle for the $\pi^0$ candidate, where the decay angle is defined as the angle between the photon momentum and the boost direction of the laboratory system in the $\pi^0$ rest frame. 
\item 
 For $\eta \to \gamma \gamma$ candidates, we use the two-photon invariant mass and the decay angle. For $\eta \to \pi^+ \pi^- \pi^0$ candidates, the $\pi^+ \pi^- \pi^0$ invariant mass is used.
\item
 For $\eta'$ candidates, we use the invariant mass of the $\pi^+ \pi^- \eta$ or $\pi^+ \pi^- \gamma$ combination and the p-value of the mass-vertex-constrained fit.
\item
 For $D$ meson candidates, we use the invariant mass of the $D$ candidate and the p-value of
 the mass-vertex-constrained fit. In three-body decays, we include the invariant masses of intermediate $\rho\,(\to\pi\pi)$,
 $K^*(\to{K}\pi)$, and $\phi(\to K^+ K^-)$ resonance candidates.
\item
 For $J/\psi$ and $D^*$ candidates, we use the invariant masses.
\item
 For $B_s^0$ meson candidates, if there are several pions or kaons in the decay, we include the invariant masses of intermediate $\rho$, $K^*$, and $a_1(\to\pi\pi\pi)$
 resonance candidates.
\item
 To suppress continuum events, we use the event-shape variable $R_2$
 (the ratio of the second to zeroth Fox-Wolfram
 moments~\cite{Fox:1978vu}), the angle between the thrust axes of the
 $B_s^0$ candidate and that of the rest of the event~\cite{Kou_2019}. All quantities are defined in the center-of-mass frame.
\end{itemize}

We train the BDT separately for each final-state particle
species and for each decay of the unstable particle. The training
result, the classifier output, is the probability ($\cal P$) that a given
candidate is signal. In addition to the variables listed above, the
training for each decay also uses the signal probabilities of all
direct decay-products.
To realize this, the training is carried out in stages, first to determine the signal probability for charged tracks, $\pi^0$, and $K_S^0$ candidates, then for $\eta$ and $J/\psi$ candidates, next for $\eta’$ candidates, then for $D$ candidates, subsequently for $D^*$ candidates, and finally for $B_s^0$ candidates.

The branching fractions of some of the $B_s^0$ decay channels used for the reconstruction have large uncertainties; two of the channels have not yet been measured. In addition, the ratio of efficiencies in data and simulation could be different in different channels. As a result, the relative contributions of various channels in simulation differ from those in data. To compensate for this difference, we introduce weights for various channels in simulation, as described in Appendix~\ref{Weights}.

We apply a requirement on the $B_s$ momentum, $|p^*(B_s) - 0.42|<0.09\gevc$, and channel-dependent requirements on ${\cal P}_{B_s}$, which are given in table~\ref{1Bs}. The above requirements are optimized to reach maximal sensitivity to the yield of $B_s-D$ pairs, as
described in the next section. The $B_s$ momentum requirement selects the dominant production channel $e^+e^- \to B_s^{*} \bar{B}_s^{*}$ with 96\% efficiency, while the channels $e^+e^- \to B_s^0 \bar{B}_s^0$ and $e^+e^- \to B_s^0 \bar{B}_s^*$,
which correspond to a total fraction of 15\%, are not included. We select $B_s^0$ candidates with invariant mass $\mbs$ in the interval ($5.25, 5.51 \gevcc$),
which is used for fitting as described below. In the case of multiple $B_s^0$ candidates, we select the one that has the highest signal probability.

\begin{table}[htbp]
		\caption{The requirements on ${\cal P}_{B_s}$ optimized for the measurement of $\Br(B_s^0 \to D/\bar{D} X)$, the
number of selected $B_s^0$ candidates, and the shift and width scaling parameters of the $\mbs$ signal functions.}
		\label{1Bs}
		\begin{center}
    \begin{tabular}{@{}lcc@{}}

  \toprule

  Decay & ${\cal P}_{B_s}$ requirement & Number of tags, $N_{B_s}$\\
  \midrule
  $B_s^0 \to D_s^{\pm} X$ & $> 0.0012$ & $12500 \pm 310$\\
   $B_s^0 \to D^0/\bar{D}^0 X$ & $> 0.0050$ & $9610 \pm 190$\\
	 $B_s^0 \to D^{\pm} X$ & $> 0.0200$ & $6485 \pm 120$\\
  \bottomrule
   \end{tabular}
		\end{center}

	\end{table}

\subsection{Selection of signal $D$ candidates}
\label{EvSelect}
To reconstruct $D$ mesons in the ROE, we use the following channels: $D_s^+ \to \phi \pi^+$, $\bar{K}^{*0} K^+$, $K_S^0 K^+$; $D^0 \to K^- \pi^+$; and $D^+ \to K^- \pi^+ \pi^+$.

Charged kaons and pions, except those from $K_S^0$ decays, are required to originate from the IP region with $dr$ $<$ 0.5 cm and $dz$ $<$ 2 cm. We require $R_{K/\pi} = L(K)/(L(K) + L(\pi)) > 0.1$ for kaons from $D_s^+$ decays and $R_{K/\pi} > 0.6$ for those from both $D^0$ and $D^+$. The requirement for pions from $D_s^+$ and $D^0$ mesons is $R_{\pi/K} = L(\pi)/(L(K) + L(\pi)) > 0.1$ and for pions from $D^+$ mesons is $R_{\pi/K} > 0.6$. 

For $\phi \to K^+ K^-$ and $K^{*0} \to K^+ \pi^-$ candidates, the invariant masses are required to be within $40 \mevcc$ and $100 \mevcc$ of the nominal $\phi$ and $K^{*0}$ masses~\cite{Workman:2022ynf}, respectively. These requirements select $\phi$ and $K^{*0}$ mesons with an efficiency of 99\% and 91\%, respectively. The $K^0_S$ candidates are reconstructed via the decay $K^0_S \to \pi^+ \pi^-$, with the selection criteria listed in ref.~\cite{PhysRevD.72.012004}, and are also required to have an invariant mass within $15\mevcc$ of the nominal $K_S^0$ mass; the efficiency of this requirement is 96\%.

To reconstruct $D_s^+$ candidates, for both $\phi$ and $K^{*0}$ resonances the selection requirement $|\cos{\theta_{\mathrm{hel}}}| > 0.3$ is applied, where the helicity angle $\theta_{\mathrm{hel}}$ is defined as the angle between the $K^-$ and $D_s^+$ momenta in the resonance rest frame. The helicity angle distributions are expected to follow a $\cos^2{\theta_{\mathrm{hel}}}$ distribution.

The requirements on the $B_s^0$ and $D$ variables described above are optimized using a two-dimensional distribution in $\mbs$ and $M(D)$ in simulation and maximizing the figure-of-merit, defined as $S/\sqrt{S + B}$, where $S$ is the number of properly reconstructed signal $B_s - D$ pairs, and $B$ is the number of all other candidates in the signal region. The average number of multiple $D$ candidates is $1.04-1.10$ for $D_s^+$ channels, 1.03 for $D^0$, and 1.17 for
$D^+$. All candidates are included in the analysis. We verify that there is no peaking behaviour in the $M(D)$ distribution from multiple $D$ candidates.

\section{Yield of $B_s^0$ tags}
\begin{figure}[htbp]
	\centering
	\includegraphics[width = 0.65\textwidth]{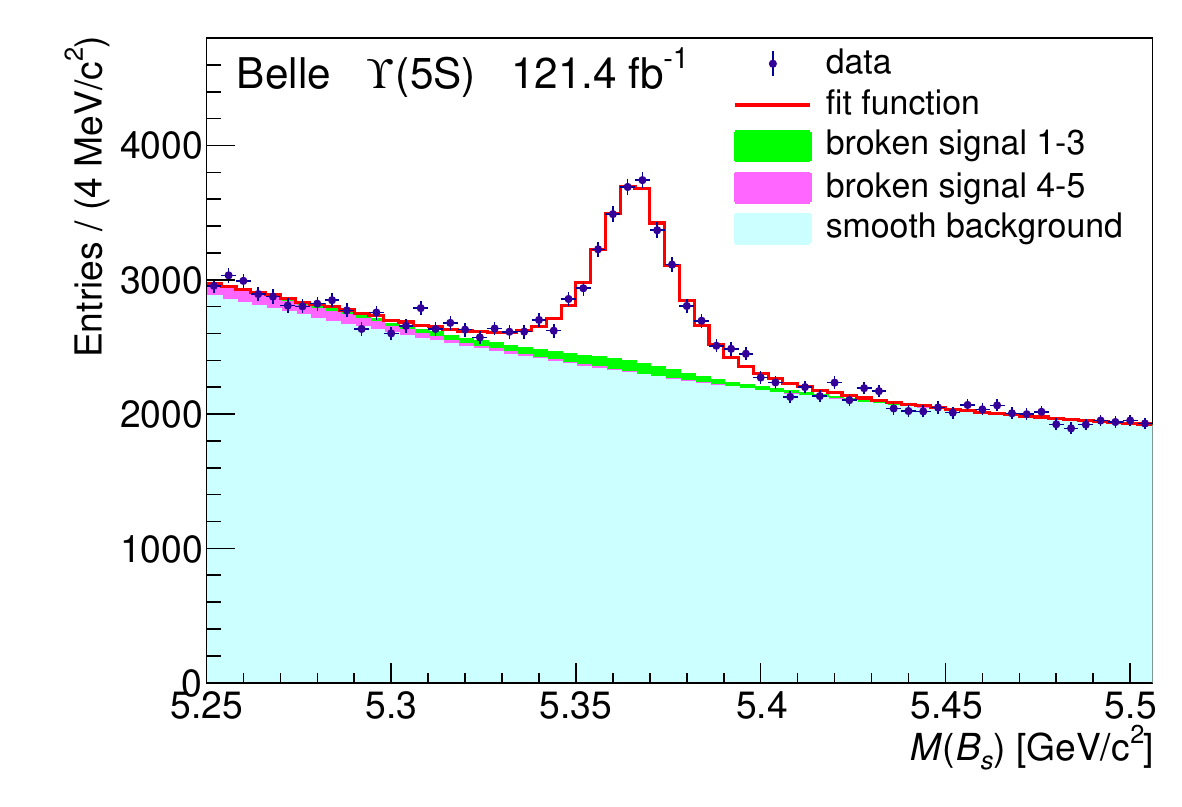}
	\caption{The mass distribution for the selected $B_s^0$ candidates. The points with error bars are data, the solid red histogram is the result of the fit, the filled green and magenta histograms are the broken-signal components, and the filled blue histogram is the smooth background component of the fit function}.\label{fig:1}
\end{figure}

The mass distribution of $B_s^0$ tag candidates, selected with the requirements optimized for the measurement of the $B_s^0 \to D_s^{\pm} X$ branching fraction, is shown in figure~\ref{fig:1}. We perform a binned likelihood fit to this distribution with a bin size of $1 \mevcc$. We fit this distribution to the sum of a correctly-reconstructed signal (CRS) component, a broken-signal component, peaking background from $B^0$, and a smooth background component. The mass distribution of the CRS events has an RMS width from 9 MeV to 32 MeV depending on the $B_s$ reconstruction channel. The determination of the CRS component shape is described below. The $B_s^0$ broken-signal components are due to signal decays with
\renewcommand\labelenumi{(\theenumi)}
\begin{enumerate}
    \item secondary interactions of final-state particles with the detector material;
    \item pions and kaons from a $D$ decay being swapped with those produced directly in a decay of a $B_s^0$;
    \item a low-momentum signal $\gamma$, $\pi^0$ or $\pi^+$ swapped with a background candidate;
    \item loss of a photon from $D_s^* \to D_s \gamma$;
    \item misidentification of a kaon as a pion.
\end{enumerate}
The shapes of these contributions are determined from simulation; their yields are fixed relative to that of the CRS. Contributions 1--3 peak in the region of CRS but have larger widths. Their yields are added to that of CRS; thus, the sum of CRS and broken-signal components 1--3 is counted as the total signal yield. The fraction of broken signal in the above sum is $9-15\%$ depending on the ${\cal P}_{B_s}$ requirement. The $B^0$ peaking background is due to Cabibbo-suppressed decays or decays with pions misidentified as kaons. The calibration of the simulation of broken-signal components 4--5 and the $B^0$ peaking background is described in Appendix~\ref{appendixPBG}. The broken signal and $B^0$ background contributions are represented in the fit as histograms. The smooth background is described by a second order polynomial.

To determine the overall shape of the CRS component, we perform the fits described above separately for each $B_s^0$ reconstruction channel. In these fits, the CRS component is described by a sum of three ($D_s^{(*)-} \pi^+ (\pi^0, \pi^+ \pi^-)$, $J/\psi K^+ K^-$) or two (other channels) Gaussian functions. We fix the relative normalizations, means, and widths of the Gaussians to the results from simulation and then introduce common parameters, representing a shift in means and broadening of the widths, which are floated to adjust for differences between data and simulation. We find good agreement between data and the fit function in each channel, which provides validation for our modelling of background. We note that the shift and the broadening factor applied to the CRS component also account for potential
mismodelling of broken-signal components 1--3 that peak in the signal region. After fitting the distributions in each channel, we add all the signal components to determine the overall signal function.

\section{Yields of $B_s^0 - D$ pairs}
\begin{figure}[htbp]
	\centering
	\includegraphics[width = 0.65\textwidth]{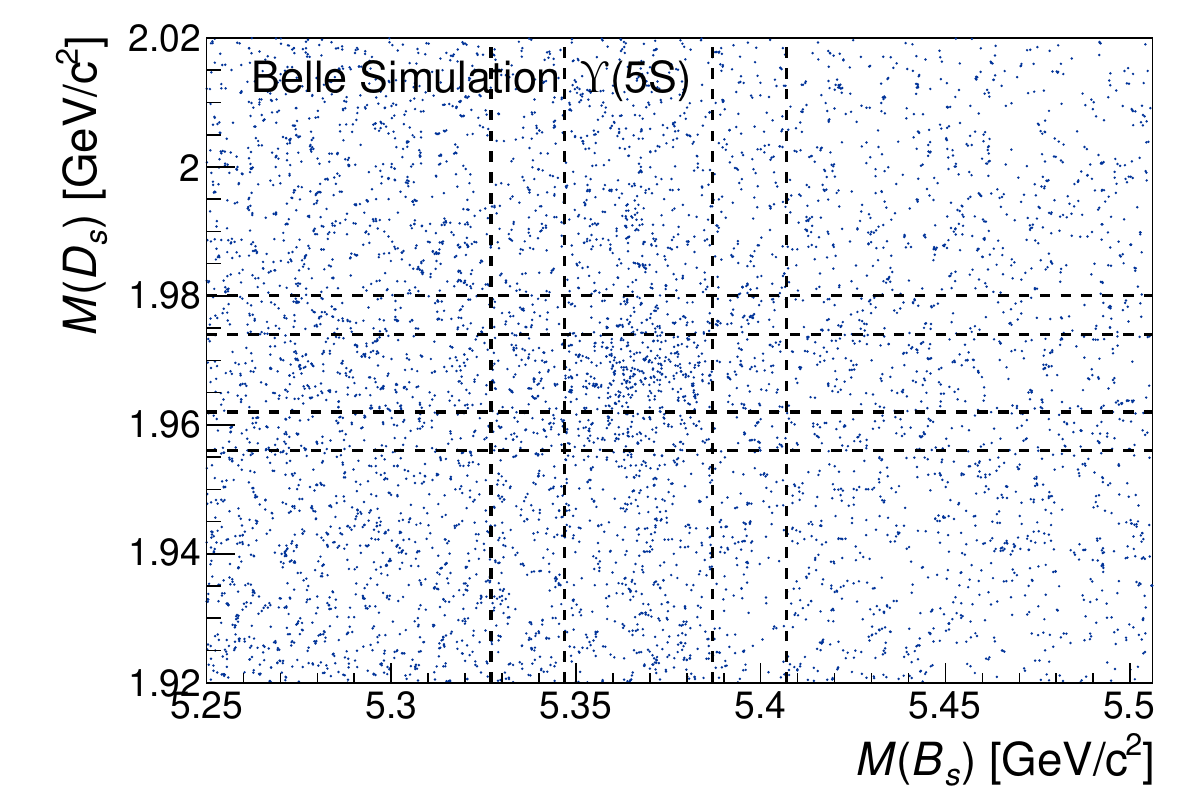}
	\caption{The distribution in $\mbs$ and $M(D_s(\to K^{*0} K))$ in simulation. Vertical and horizontal dashed lines indicate $\mbs$ and $\mds$ sideband and signal regions.\label{fig:2}}
\end{figure}

To obtain the number of $B_s - D$ pairs, we perform a two-dimensional binned likelihood fit to the distribution in $\mbs$ and $M(D)$, with a bin size in $M(B_s)$ of $1 \mevcc$ and in $M(D)$ of $0.5 \mevcc$. The distribution for $D_s^+ \to \bar{K}^{*0} K^+$ in simulation is shown in figure~\ref{fig:2}. The boundaries of signal regions and sidebands are listed in table~\ref{SBs}.

\begin{table}[htbp]
		\caption{The boundaries of the signal region and sidebands along each axis in the two-dimensional distributions.}
		\label{SBs}
		\begin{center}
    \begin{tabular}{@{}lcccc@{}}

  \toprule

  Region & $\mbs$ & $\mds$ & $\mdz$ & $\mdp$ \\
  \midrule
  Signal region & (5.347, 5.387) & (1.962, 1.974) & (1.849, 1.869) & (1.854, 1.874)\\
  Left sideband & (5.250, 5.327) & (1.920, 1.956) & (1.800, 1.839) & (1.805, 1.844)\\
  Right sideband & (5.407, 5.510) & (1.980, 2.020) & (1.879, 1.935) & (1.884, 1.940)\\
  \bottomrule
	 
   \end{tabular}
		\end{center}

	\end{table}

Our fit function has four components: both $B_s$ and $D$ candidates are
signal (SS), the $B_s$ candidate is signal and the $D$ candidate is background
(SB), the $B_s$ candidate is background and the $D$ candidate is signal (BS), and
both candidates are background (BB), each being the product of one-dimensional signal or background mass functions. The $B_s$ signal component is a sum of all peaking contributions: CRS, broken signal and $B^0$ peaking background. The shapes of all these contributions and their relative yields are the same as in figure~\ref{fig:1}. For the SS component, we fix the absolute yield of the $B^0$ peaking background taking into account the inclusive branching fraction $\Br(B^0 \to D/\bar{D} X)$. The $M(D)$ signal function is described by a sum of three Gaussians. Its shift and width scaling factors are determined from a fit to the one-dimensional $M(D)$ distribution for inclusively produced $D$ mesons that satisfy $p^*(D_s)<2.7\gevc$, $p^*(D^0) < 2.5\gevc$, and $p^*(D^+) < 2.4\gevc$. These requirements have close to 100\% efficiency for $D$ mesons produced in $B_s^0$ decays. The values of the shifts and width scaling  factors for various $D$ decay channels are listed in table~\ref{1Ds}; they are fixed in the two-dimensional fits. No peaking background is observed in the $M(D)$ distribution.

\begin{table}[htbp]
 \caption{The shift and width scaling parameters of the $M(D)$ signal functions.}
 \label{1Ds}
 \begin{center}
    \begin{tabular}{@{}lrc@{}} \toprule
  Channel & \multicolumn{1}{c}{Shift, $\mathrm{MeV}/c^2$} & Scaling\\
  \midrule
  $D_s^+ \rightarrow \phi \pi^+$ & $-0.21 \pm 0.02$ & $0.978 \pm 0.005$\\
  $D_s^+ \rightarrow \bar{K}^{*0} K^+$ & $-0.24 \pm 0.05$ & $1.004 \pm 0.017$\\
  $D_s^+ \rightarrow K_S^0 K^+$ & $-0.25 \pm 0.06$ & $1.059 \pm 0.022$\\
  $D^0 \rightarrow K^- \pi^+$ & $0.16 \pm 0.01$ & $0.982 \pm 0.002$\\
  $D^+ \rightarrow K^- \pi^+ \pi^+$ & $0.08 \pm 0.02$ & $0.968 \pm 0.003$\\
  \bottomrule
 \end{tabular}
  \end{center}
\end{table}

In the BS component the smooth background is described by an exponential function ($D_s^+$ or $D^+$ in the ROE) or a constant ($D^+$ in the ROE), and in the BB component by a first order polynomial. The background dependence in $\mds$ in the SB component is constant ($\bar{K}^{*0} K^+$, $K_S^0 K^+$) or linear ($\phi \pi^+$), and in the BB component it is linear for all channels. The background $\mdz$ and $\mdp$ functions in the SB and BB components are linear. The parameters of the background functions are free in the fit.

\begin{figure}[htbp]
\centering
\includegraphics[width=1.0\textwidth]{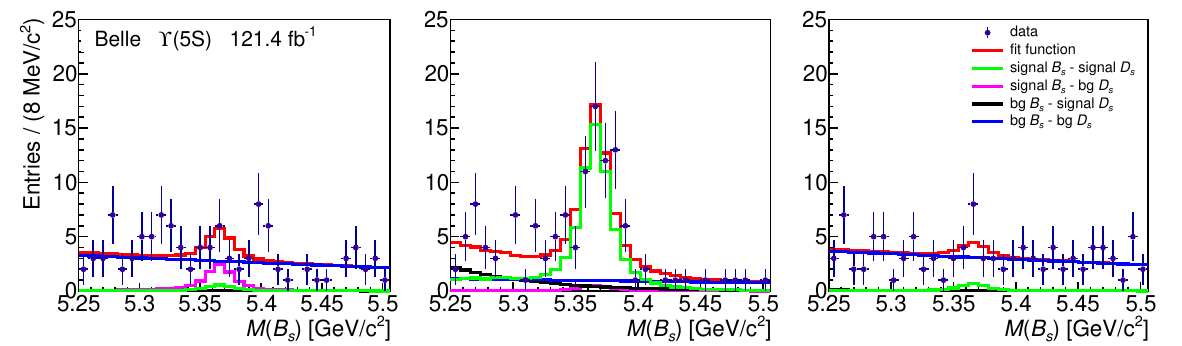}
\includegraphics[width=1.0\textwidth]{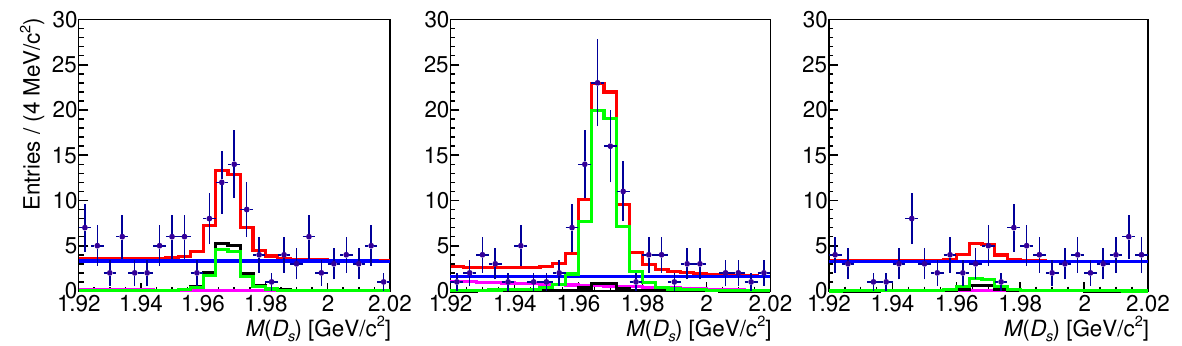}
\caption{Projections of the two-dimensional fit to the distribution in $\mbs$ and $\mds$ for the $D_s^+ \to \phi \pi^+$ channel onto $\mbs$ (top) and $\mds$ (bottom) axis. The left and right figures correspond to the projections in the left and right sideband regions, and the central figures show projections in the signal region. Blue points with error bars represent the data. The solid red histograms show the total fit function, while the solid green, black, magenta, and blue histograms show the SS, BS, SB and BB components, respectively. \label{fig:3}}
\end{figure}

\begin{figure}[htbp]
\centering
\includegraphics[width=1.0\textwidth]{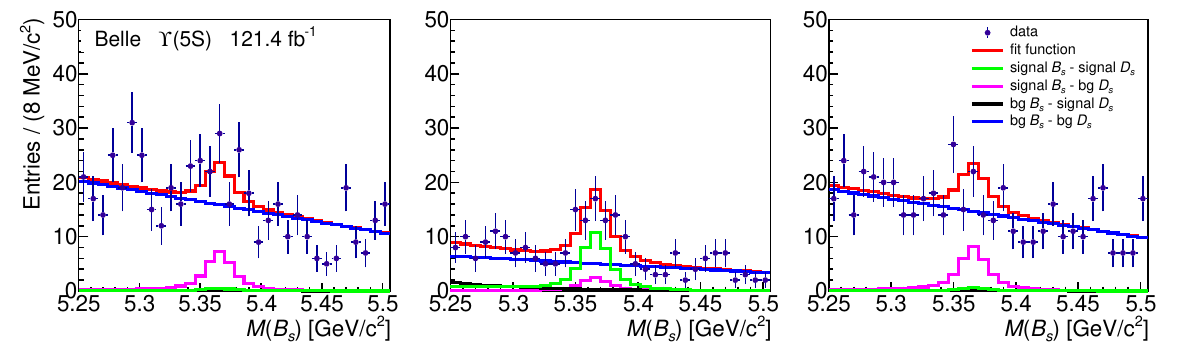}
\includegraphics[width=1.0\textwidth]{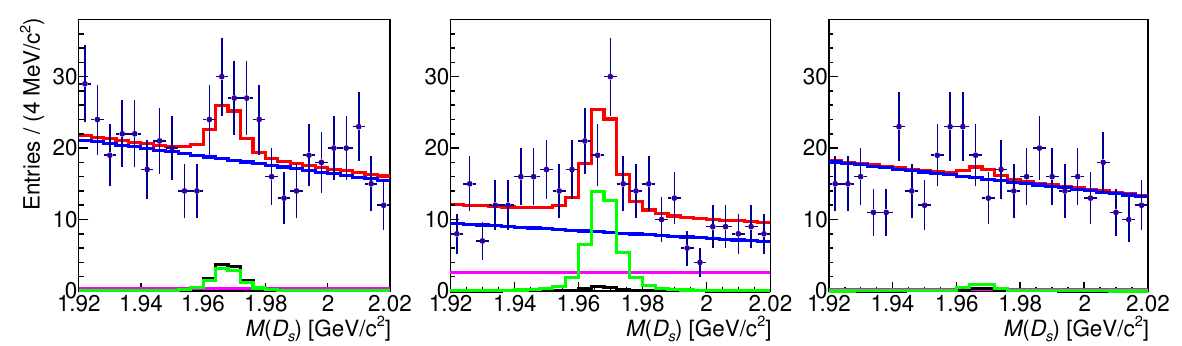}
\caption{Projections of the two-dimensional fit to the distribution in $\mbs$ and $\mds$ for the $D_s^+ \to \bar{K}^{*0} K^+$ channel onto $\mbs$ (top) and $\mds$ (bottom) axis. The left and right figures correspond to the projections in the left and right sideband regions, and the central figures show projections in the signal region. The legend is the same as in Fig.~\ref{fig:3}.\label{fig:4}}
\end{figure}

\begin{figure}[htbp]
\centering
\includegraphics[width=1.0\textwidth]{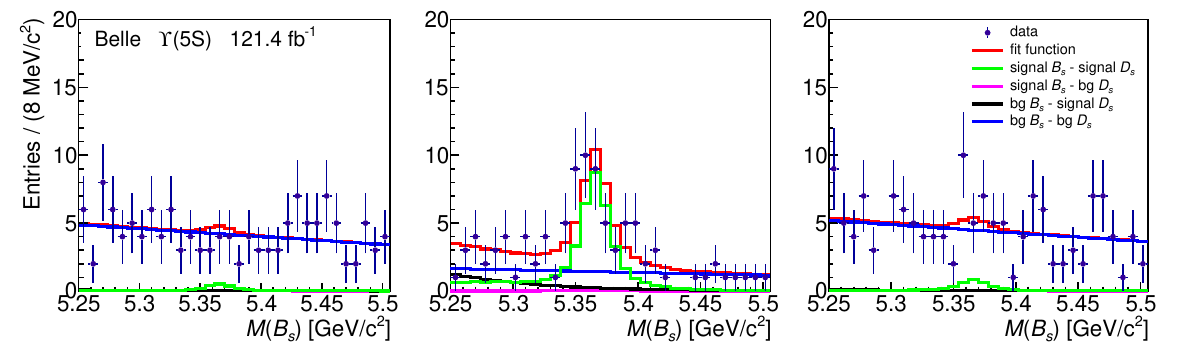}
\includegraphics[width=1.0\textwidth]{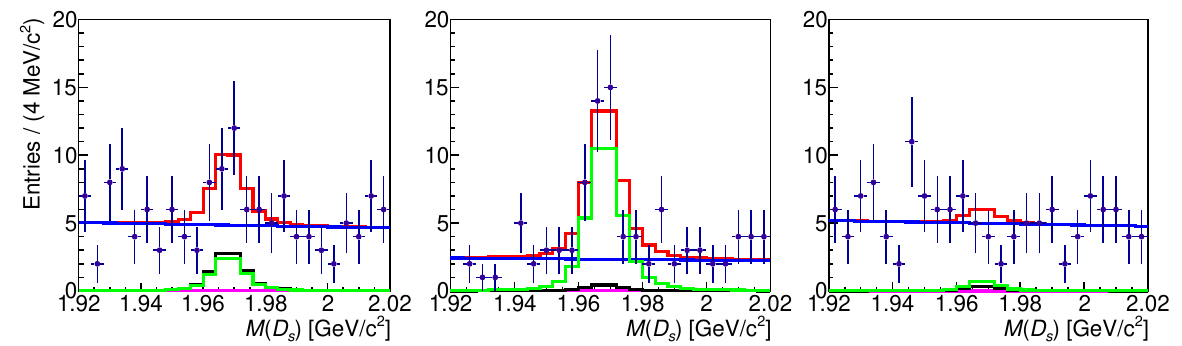}
\caption{Projections of the two-dimensional fit to the distribution in $\mbs$ and $\mds$ for the $D_s^+ \to K_S^0 K^+$ channel onto $\mbs$ (top) and $\mds$ (bottom) axis. The left and right figures correspond to the projections in the left and right sideband regions, and the central figures show projections in the signal region. \label{fig:5}}
\end{figure}

\begin{figure}[htbp]
\centering
\includegraphics[width=1.0\textwidth]{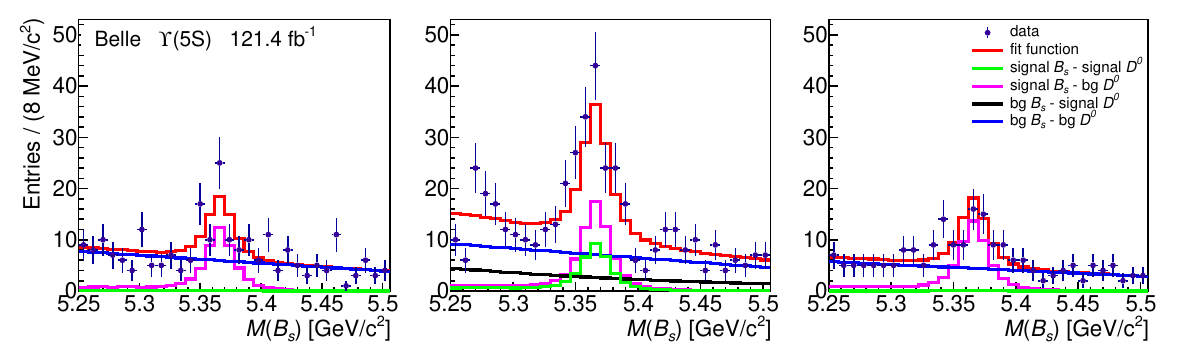}
\includegraphics[width=1.0\textwidth]{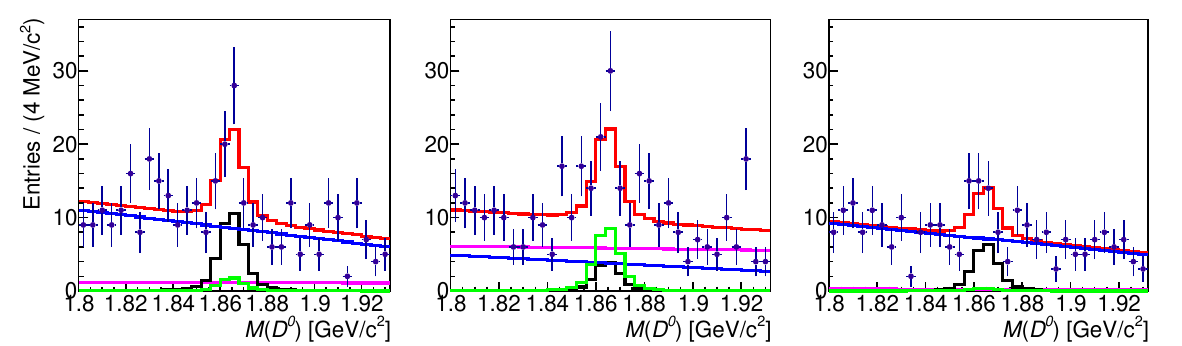}
\caption{Projections of the two-dimensional fit to the distribution in $\mbs$ and $\mdz$ onto $\mbs$ (top) and $\mdz$ (bottom) axis. The left and right figures correspond to the projections in the left and right sideband regions, and the central figures show projections in the signal
region.\label{fig:6}}
\end{figure}

\begin{figure}[htbp]
\centering
\includegraphics[width=1.0\textwidth]{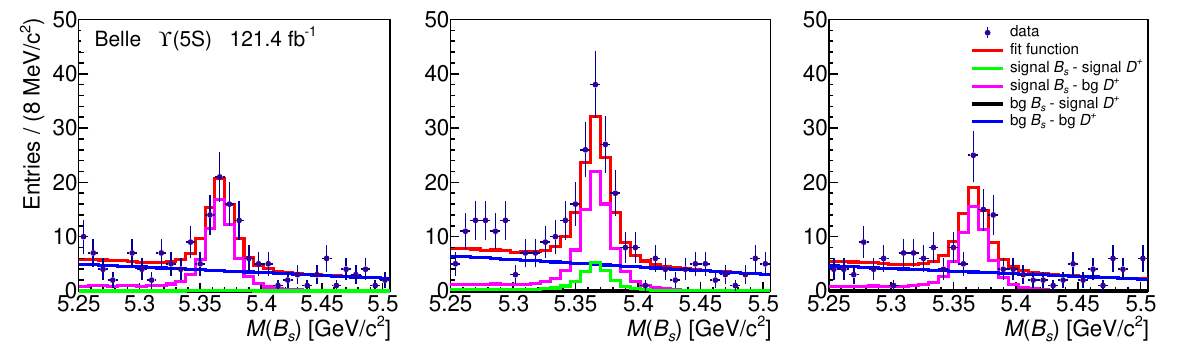}
\includegraphics[width=1.0\textwidth]{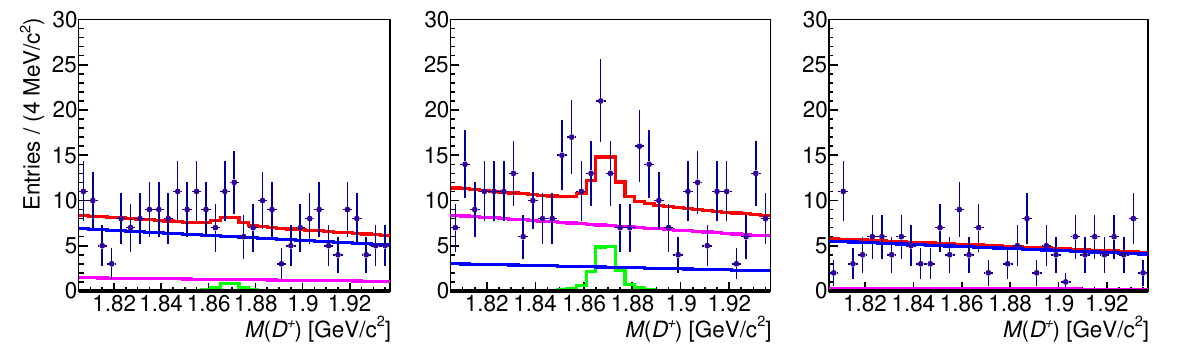}
\caption{Projections of the two-dimensional fit to the distribution $\mbs$ and $\mdp$ onto $\mbs$ (top) and $\mdp$ (bottom) axis. The left and right figures correspond to the projections in the left and right sideband regions, and the central figures show projections in the signal
region.\label{fig:7}}
\end{figure}

In the two-dimensional $\mbs$ and $\mds$ distributions, the ratio of yields for two components, SS and BS, should not depend on the reconstructed channel of $D_s^+$. Moreover, it is expected that the shape of the $\mbs$ background in the BS component does not depend on the $D_s^+$ channel. We thus perform a simultaneous two-dimensional fit to the distributions in $\mbs$ and $\mds$ for all three $D_s^+$ channels with $N_{BS}/N_{SS}$ and the slope of the exponential function in the BS components being common free parameters. Projections of the fit result on each of the axes in the signal and sideband regions defined in table~\ref{SBs} are shown in figures~\ref{fig:3}--\ref{fig:7}. The fit results for yields are given in table~\ref{tab2D}.

\begin{table}[htbp]
 \caption{The yields of the two-dimensional fit, the branching fractions of $D$ mesons, the reconstruction efficiency, and calculated $B_s^0$ branching fractions.}
 \label{tab2D}
 \begin{center}
    \begin{tabular}{@{}lcccc@{}} \toprule
  Decay & $N_{B_s-D}$ & $\Br_D, \%$ & $\varepsilon_{D}^{\mathrm{ROE}}, \%$ & $\Br(B_s^0 \to D/\bar{D} X), \%$\\
  \midrule
  $B_s^0 \to D_s^{\pm} X$\\
  \tabitem $\phi \pi^+$ & 85 $\pm$ 12 & 5.37 $\pm$ 0.10 & 17.3 $\pm$ 0.8 & 73.0 $\pm$ 10.6 $\pm$ 5.2 \\
  \tabitem $\bar{K}^{*0} K^+$ & 63 $\pm$ 13 & 5.37 $\pm$ 0.10 & 17.3 $\pm$ 0.8 & 54.1 $\pm$ 11.7 $\pm$ 3.7\\
  \tabitem $K_S^0 K^+$ & 55 $\pm$ 10 & 1.450 $\pm$ 0.035 & 34.4 $\pm$ 1.9 & 88.2 $\pm$ 16.2 $\pm$ 7.0\\
  \midrule
  $B_s^0 \to D^0/\bar{D}^{0} X$ & 56 $\pm$ 16 & 3.947 $\pm$ 0.030 & 68.2 $\pm$ 5.1 & 21.5 $\pm$ 6.1 $\pm$ 1.8 \\
  $B_s^0 \to D^{\pm} X$ & 34 $\pm$ 12 & 9.38 $\pm$ 0.16 & 44.4 $\pm$ 4.0 & 12.6 $\pm$ 4.6 $\pm$ 1.3\\
  \bottomrule
 \end{tabular}
  \end{center}
\end{table}

\section{Branching fractions and systematic uncertainties}
\label{uncert_sec}
We calculate the branching fractions using Eq.~\eqref{Eq1}, where the yields of $B_s^0$ tags and $B_s - D$ pairs, the reconstruction efficiency for $D$ mesons, and the $D$ mesons branching fractions are given in table~\ref{tab2D}.

We determine the reconstruction efficiency for $D$ mesons in the ROE using simulation. 
In the calculation of the reconstruction efficiency for the $D_s^+ \to \phi \pi^+$ and $D_s^+ \to \bar{K}^{*0} K^+$ channels, we use the $D_s^+ \to K^+ K^- \pi^+$ branching fraction. Thus, the efficiency $\varepsilon_{D_s}^{\mathrm{ROE}}$ includes the selection efficiency for the corresponding $K^+ K^- \pi^+$ Dalitz plot regions.

The values obtained for $\Br(B_s^0 \to D/\bar{D} X)$ are listed in table~\ref{tab2D}, where the first uncertainty is statistical and the second is systematic. The statistical uncertainty includes contributions from $N_{B_s}$ and $N_{B_s-D}$.

Sources of systematic uncertainties are listed in tables~\ref{uncert} and~\ref{uncertD0} and are described below.
\begin{itemize}
  \item To account for the uncertainty on the $M(B_s)$ signal shape, we vary the
shift and width scaling parameters of the CRS component by $\pm 1\sigma$ in each
reconstruction channel individually. Corresponding deviations in the measured branching fractions are added in quadrature for all channels to obtain the total uncertainty due to the description of the $M(B_s)$ signal shape. The uncertainty due to the description of the $M(D)$ signal shape is negligible.

  \item Variation of CRS shape parameters described in the previous item partly
accounts for potential mismodelling of the broken-signal components that peak in the signal region. In addition, we multiply each of the above components by factors of 0.75 or 1.25 simultaneously in all channels. The broken-signal components in the left $M(B_s)$ sideband are calibrated as described in Appendix~\ref{appendixPBG}; we vary the corresponding yields by $\pm 1\sigma$. Deviations in the measured branching fractions due to all variations are added in quadrature to obtain the total uncertainty due to possible mismodelling of the broken signal. The uncertainty due to $B^0$ peaking background is negligible.

  \item The uncertainty on the smooth background shape in the two-dimensional fit is obtained by varying its shape parameters: we change all constant and linear functions into exponential functions, and also add a cross term that does not arise as a result of multiplying one-dimensional functions. The deviations in the fit results are added in quadrature, and the sum is assigned as the systematic uncertainty.
  
  \item The systematic uncertainty from the track reconstruction efficiency, estimated using partially reconstructed $D^{*+} \to D^0 \pi^+$, $D^0 \to \pi^+ \pi^- K_S^0$ and $K_S^0 \to \pi^+ \pi^-$ events~\cite{BaBar:2014omp}, is 0.35\% per track. We take 1.1\% as the associated systematic uncertainty for the $D_s^+$ and $D^+$ channels, and 0.7\% for the $D^0$ channel.
  
  \item The uncertainty from the $K/\pi$ identification efficiency due to a possible difference between MC and data is studied using $D^{*+} \to D^0 (K^- \pi^+) \pi^+$ decays~\cite{BaBar:2014omp}. The uncertainty is 2.1\% for $D_s^+ \to \phi \pi^+$, 1.9\% for $D_s^+ \to \bar{K}^{*0} K^+$, and 0.7\% for $D_s^+ \to K_S^0 K^+$; 1.2\% for $D^0 \to K^- \pi^+$ and 3.0\% for $D^+ \to K^- \pi^+ \pi^+$.
  
  \item The uncertainty from the $K_S^0$ reconstruction efficiency, which is studied using $D^{*+} \to D^0 (\pi^+ \pi^- K_S^0) \pi^+$ decays~\cite{BaBar:2014omp}, is found to be 2.3\%.
  
  \item We account for the uncertainty due to the difference between the $D_s^+ \to K^+ K^- \pi^+$ Dalitz plot in simulation and data. This difference is studied for inclusive $D_s^+$ mesons, and the following correction factors for the $\phi$ and $K^{*0}$ reconstruction efficiencies are applied
  \begin{equation}
    \begin{aligned}
      r_{\phi} = {\varepsilon_{\phi}^{\mathrm{data}}}/{\varepsilon_{\phi}^{\mathrm{MC}}} = 0.947 \pm 0.008; \\
      r_{K^*} = \varepsilon_{K^{*}}^{\mathrm{data}}/{\varepsilon_{K^*}^{\mathrm{MC}}} = 1.042 \pm 0.008.
    \end{aligned}
  \end{equation}
  The statistical uncertainty for this factor is included as a systematic uncertainty, which is estimated to be 0.8\%.

  \item Since the momentum spectra of $D$ mesons from $B_s^0$ decays may differ between simulation and data, we estimate the uncertainty due to the dependence of reconstruction efficiency on the momentum. We examine $D_s^+$ mesons in the ROE and measure their yields for center-of-mass momenta below and above $1.5 \gevc$. The yield ratio in data, $1.15 \pm 0.22$, is in good agreement with that in simulation, 1.17. We introduce weights for simulated events according to the above uncertainty in data and include the corresponding deviations in the efficiency as systematic uncertainties. We find $0.2-0.8\%$ for $D_s^+$, 0.2\% for $D^0$, and below 0.1\% for $D^+$.

 \item The reconstruction efficiency of the $B_s^0$ tag depends on multiplicity in the ROE, primarily because of the requirement of one $B_s^0$ candidate per event. If there is a reconstructed $D$ in the ROE, then the multiplicity is lower than the average value and the $B_s^0$ reconstruction efficiency is expected to be slightly higher. According to simulation, the effect is at the 5\% level. To estimate the corresponding systematic uncertainty, we measure the inclusive branching fractions of $B^+$ and $B^0$ mesons using part of the $\Upsilon(4S)$ data with an integrated luminosity of 571 fb$^{-1}$ corresponding to the same inner detector configuration as the $\Upsilon(5S)$ data. The analysis procedure is the same as for $B_s^0$ mesons. The $B^+$ and $B^0$ decay channels used in the FEI are listed in table~\ref{tab3} in Appendix~\ref{appendixA}. We find good agreement with the previous Belle measurement~\cite{Belle:2023yfw} and assign a systematic uncertainty of 3.6\% for $D_s^+$ and  2.6\% for $D^0$ and $D^+$ based on the uncertainty of the results.
 
  \item The contributions from the limited size of the MC samples are estimated to be $4.4\%-5.7\%$ for $D_s^+$, 7.5\% for $D^0$, and 9.0\% for $D^+$. 

  \item The uncertainty on the world averages $\Br(D_s^+ \to K^+ K^- \pi^+ )$, $\Br(D_s^+ \to K_S^0 K^+)$, and $\Br(D^0 \to K^- \pi^+)$ are 1.9\%, 2.5\%, and 1.7\%, respectively~\cite{Workman:2022ynf}.
  
\end{itemize}

{The total systematic uncertainty is obtained by adding the various contributions
in quadrature.}

\begin{table}[htbp]
		\caption{Systematic uncertainties in the measurement of ${\cal B}(B_s^0 \rightarrow D_s^{\pm} X)$ (in \%).}
		\label{uncert}
		\begin{center}
 \begin{tabular}{@{}lcccc@{}}

   \toprule
    & \multicolumn{3}{c}{Channel}\\
\raisebox{1.5ex}[0cm][0cm]{Source}
& $\phi \pi^{+}$ & $\bar {K}^{*0} K^+$ & $K_S^0 K^+$ & \raisebox{1.5ex}[0cm][0cm]{Combined} \\
    \midrule
    Signal shape & 2.3 & 1.8 & 1.6 & 2.0 \\
    Broken signal & 0.9 & 0.9 & 0.9 & 0.9 \\
    Smooth background &  1.6 &  1.0 &  1.1 &  1.4\\
    \midrule
    Tracking & 1.1 & 1.1 & 1.1 & 1.1 \\
    $K/\pi$ identification & 2.1 & 1.9 & 0.7 & 1.7\\
    $K^0_S$ reconstruction & -- & -- & 2.3 & 0.6\\
    $D_s$ momentum & 0.8 & 0.6 & 0.2 & 0.6 \\
    Dalitz plot & 0.8 & 0.8 & -- & 0.6\\
    FEI efficiency & 3.6 & 3.6 & 3.6 & 3.6\\
    MC statistics & 4.4 & 4.5 & 5.7 & 2.7\\
	\midrule
	 ${\cal B}(D_s \to K K \pi)$ & 1.9 & 1.9 & -- & 1.4\\
	 ${\cal B}(D_s \to K_S K)$ & -- & -- & 2.4 & 0.6 \\
	 ${\cal B}(K_S^0 \to \pi^+ \pi^-)$ & -- & -- & $<$ 0.1 & --\\
	 \midrule
	 Total & 7.2 & 6.9 & 7.9 & 5.9\\
	 \bottomrule
	 
			\end{tabular}
		\end{center}

	\end{table}

\begin{table}[htbp]
		\caption{Systematic uncertainties in the measurements of ${\cal B}(B_s^0 \to D^0/\bar{D}^0 X)$ and ${\cal B}(B_s^0 \rightarrow D^{\pm} X)$ (in \%).}
		\label{uncertD0}
		\begin{center}
    \begin{tabular}{@{}lcc@{}}

  \toprule

  Source & $B_s^0 \to D^0/\bar{D}^0 X$ & $B_s^0 \to D^{\pm} X$ \\
  \midrule
  Signal shape & 2.0 & 0.6 \\
  Broken signal & 1.1 & 2.9 \\
  Smooth background &  0.3 &  0.9\\
  \midrule
  Tracking & 0.7 & 1.1\\
  $K/\pi$ identification & 1.2 & 3.0\\
  $D$ momentum & 0.2 & $<$0.1\\
  FEI efficiency & 2.6 & 2.6\\
  MC statistics & 7.5 & 9.0\\

	\midrule
	 ${\cal B}(D \to K \pi (\pi))$ & 0.8 & 1.7\\
   \midrule
	 Total & 8.4 & 10.5\\
	 \bottomrule
	 
			\end{tabular}
		\end{center}

	\end{table}

\section{Average branching fractions and $f_s$}
\label{sumBF}
Averaging the $\Br(B_s^0 \to D_s^{\pm}X)$ over three $D_s^+$ channels, we obtain
\begin{equation}
  \Br(B_s^0 \to D_s^{\pm}X) = (68.6 \pm 7.2 \pm 4.0)\%,
\end{equation}
where the correlations in statistical uncertainty on the number of $B_s^0$ tags and systematic
uncertainties shown in table~\ref{uncert} are taken into account using the fitting method described in ref.~\cite{PhysRevD.107.052008}. The p-value of this fit is 28\%.

The measurement of $\Br(B_s^0 \to D_s^{\pm}X)$ is in agreement with the previous Belle measurement using semileptonic tagging, (60.2 $\pm$ 5.8 $\pm$ 2.3)\%~\cite{Belle:2021qxu}, which when rescaled with the most recent values of $D_s^+$ branching fractions gives $(60.5 \pm 5.8 \pm 2.2)\%$. Averaging our measurement with the previous result after rescaling, and taking into account correlated uncertainties as shown in table~\ref{uncert_new_old}, gives
\begin{equation}
	{\cal B}(B_s^0 \to D_s^{\pm} X) = (63.4 \pm 4.5 \pm 2.2)\%.
  \label{DsX_av}
\end{equation}
\begin{table}[htbp]
		\caption{Systematic uncertainties in the ${\cal B}(B_s^0 \rightarrow D_s^{\pm} X)$ measurement in ref.~\cite{Belle:2021qxu} and in this work (in \%).}
		\label{uncert_new_old}
		\begin{center}
 \begin{tabular}{@{}lccc@{}}

  \toprule
  Source & Ref.~\cite{Belle:2021qxu} & This work & Combined \\
  \midrule
  Uncorrelated & 3.0 & 5.3 & 2.6 \\
  \midrule
  Tracking & 1.1 & 1.1 & 1.1 \\
  $K/\pi$ identification & 1.3 & 1.7 & 1.5\\
	 ${\cal B}(D_s \to K K \pi)$ & 1.5 & 1.4 & 1.4\\
	 ${\cal B}(D_s \to K_S K)$ & 0.4 & 0.6 & 0.5 \\
	 \midrule
	 Total & & & 3.5\\
	 \bottomrule
	 
			\end{tabular}
		\end{center}

	\end{table}

{Using this value and the ratio $\Br(B_s^0 \to D^0/\bar{D}^0X)/\Br(B_s^0 \to D_s^{\pm} X) = 0.416 \pm 0.018 \pm 0.092$ measured in ref.~\cite{Belle:2023yfw}, we obtain
\begin{equation}
  \Br(B_s^0 \to D^0/\bar{D}^0 X) = \Br(B_s^0 \to D_s^{\pm} X) \, \frac{\Br(B_s^0 \to D^0/\bar{D}^0 X)}{\Br(B_s^0 \to D_s^{\pm} X)} = (26.4 \pm 2.2 \pm 5.9)\%,
\end{equation}
where the uncertainty is similar to that of the direct measurement shown in table~\ref{tab2D}. We average the two values taking into account their
correlation to obtain
\begin{equation}
   \Br(B_s^0 \to D^0/\bar{D}^0 X) = (23.9 \pm 4.1 \pm 1.8)\%.
\end{equation}
Systematic uncertainties for the direct measurement, the calculated value, and their average are presented in table~\ref{D0_uncert}.
}

 \begin{table}[htbp]
		\caption{Systematic uncertainties in the direct ${\cal B}(B_s^0 \rightarrow D^0/\bar{D}^0 X)$ measurement and in the value of the product $\Br(B_s^0 \to D_s^{\pm} X) \times \frac{\Br(B_s^0 \to D^0/\bar{D}^0 X)}{\Br(B_s^0 \to D_s^{\pm} X)}$ (in \%).}
		\label{D0_uncert}
		\begin{center}
\begin{tabular}{@{}lccc@{}}

  \toprule
  Source & Direct measurement 
 & $\Br(B_s^0 \to D_s^{\pm} X) \, \frac{\Br(B_s^0 \to D^0/\bar{D}^0 X)}{\Br(B_s^0 \to D_s^{\pm} X)}$ & Combined \\
  \midrule
  Uncorrelated & 7.5 & 22.4 & 7.1 \\
  \midrule
  Signal shape & 2.0 & 0.8 & 1.3 \\
  Broken signal & 1.1 & 0.3 & 0.6\\
  Tracking & 0.7 & 0.7 & 0.7 \\
  $K/\pi$ identification & 1.2 & 1.6 & 1.4\\
  $D$ momentum & 0.2 & 0.5 & 0.4\\
  FEI efficiency & 2.6 & 1.4 & 1.9 \\
  ${\cal B}(D^0 \to K^- \pi^+)$ & 0.8 & 0.8 & 0.8 \\
	 \midrule
	 Total & & & 7.7\\
	 \bottomrule
	 
			\end{tabular}
		\end{center}

	\end{table}

The sum of three branching fractions, $\Br(B_s^0 \to D_s^{\pm}X)$,  $\Br(B_s^0 \to D^0/\bar{D}^0X)$, and $\Br(B_s^0 \to D^{\pm}X)$, is equal to $(99.9 \pm 7.6 \pm 3.8)\%$, where the correlations in systematic uncertainties are taken into account. The corresponding sums for $B^+$ and $B^0$ are $(109.0 \pm 4.5)\%$ and $(106.5 \pm 5.2)\%$, respectively~\cite{Workman:2022ynf}. These sums are not expected to depend upon the flavor of the spectator quark. Thus, averaging the results for $B^+$ and $B^0$, we find $(107.9 \pm 3.4)\%$; the sum for $B_s^0$ agrees with this value.

Using the average $\Br(B_s^0 \to D_s^{\pm}X)$ from Eq.~\eqref{DsX_av}, we recalculate $f_s$ in~\cite{Belle:2023yfw}
\begin{equation}
	f_s = (21.8 \pm 0.2 \pm 2.0)\%.
\end{equation}

To improve the accuracy of $f_s$, the relation
\begin{equation}
  f_s + f_{B\bar{B}X} + f_{\cancel{B}} = 1
  \label{constraint}
\end{equation}
is used, where $f_{B\bar{B}X} = (75.1 \pm 4.0)\%$~\cite{Belle:2021lzm} is the production rate of $B\bar{B}X$ events at the $\Upsilon(5S)$ and $f_{\cancel{B}}$ is the production rate of $b \bar{b}$ events without open-bottom mesons in final states; the contribution of known channels is $f_{\cancel{B}}^{known}$ = (4.9 $\pm$ 0.6)\%~\cite{Belle:2021lzm}. We fit to three measurements, $f_s$, $f_{B\bar{B}X}$ and $f_{\cancel{B}}$, applying (\ref{constraint}) as a constraint. We obtain
\begin{equation}
 \begin{aligned} 
f_s = (21.4^{+1.5}_{-1.7})\%;\\
f_{B\bar{B}X} = (73.8^{+1.5}_{-2.9})\%;\\
f_{\cancel{B}} = (4.8^{+3.6}_{-0.5})\%.
 \end{aligned}
\end{equation}
This result for $f_s$ supersedes the previous value of the $B_s^0$ production rate $f_s = (22.0^{+2.0}_{-2.1})\%$~\cite{Belle:2023yfw}.

\section{Conclusions}
To conclude, we have measured the inclusive branching fractions for $B_s^0$ decays into $D$ mesons, using full hadronic reconstruction of one $B_s^0$ in $e^+e^- \to B_s^{*} \bar{B}_s^{*}$. We find
\begin{equation}
  \begin{aligned}
    {\cal B}(B_s^0 \rightarrow D_s^{\pm} X) = (68.6 \pm 7.2 \pm 4.0)\%, \\
	{\cal B}(B_s^0 \to D^0/\bar{D}^0 X) = (21.5 \pm 6.1 \pm 1.8)\%, \\
	{\cal B}(B_s^0 \rightarrow D^{\pm} X) = (12.6 \pm 4.6 \pm 1.3)\%.
  \end{aligned}
  \label{Results}
\end{equation}
We improve the accuracy of $\Br(B_s^0 \to D_s^{\pm}X)$ by averaging with the result of previous measurement~\cite{Belle:2021qxu} and obtain
\begin{equation}
	{\cal B}(B_s^0 \to D_s^{\pm} X) = (63.4 \pm 4.5 \pm 2.2)\%.
\end{equation}
Multiplying this value by the ratio $\Br(B_s^0 \to D^0/\bar{D}^0 X)/{\Br(B_s^0 \to D_s^{\pm} X)}$ measured in ref.~\cite{Belle:2023yfw} and averaging the result obtained for $\Br(B_s^0 \to D^0/\bar{D}^0 X)$ with the direct measurement presented in Eq.~\eqref{Results}, we find
\begin{equation}
   \Br(B_s^0 \to D^0/\bar{D}^0 X) = (23.9 \pm 4.1 \pm 1.8)\%.
\end{equation}
Using the average value of $\Br(B_s^0 \to D_s^{\pm} X)$, we update the production fractions
\begin{equation}
 \begin{aligned} 
f_s = (21.4^{+1.5}_{-1.7})\%;\\
f_{B\bar{B}X} = (73.8^{+1.5}_{-2.9})\%;\\
f_{\cancel{B}} = (4.8^{+3.6}_{-0.5})\%.
 \end{aligned}
\end{equation}
These results supersede those reported in refs.~\cite{Belle:2023yfw,Belle:2021lzm}.

\section*{Acknowledgements}
This work, based on data collected using the Belle detector, which was
operated until June 2010, was supported by 
the Ministry of Education, Culture, Sports, Science, and
Technology (MEXT) of Japan, the Japan Society for the 
Promotion of Science (JSPS), and the Tau-Lepton Physics 
Research Center of Nagoya University; 
the Australian Research Council including grants
DP210101900, 
DP210102831, 
DE220100462, 
LE210100098, 
LE230100085; 
Austrian Federal Ministry of Education, Science and Research (FWF) and
FWF Austrian Science Fund No.~P~31361-N36;
National Key R\&D Program of China under Contract No.~2022YFA1601903,
National Natural Science Foundation of China and research grants
No.~11575017,
No.~11761141009, 
No.~11705209, 
No.~11975076, 
No.~12135005, 
No.~12150004, 
No.~12161141008, 
and
No.~12175041, 
and Shandong Provincial Natural Science Foundation Project ZR2022JQ02;
the Czech Science Foundation Grant No. 22-18469S;
Horizon 2020 ERC Advanced Grant No.~884719 and ERC Starting Grant No.~947006 ``InterLeptons'' (European Union);
the Carl Zeiss Foundation, the Deutsche Forschungsgemeinschaft, the
Excellence Cluster Universe, and the VolkswagenStiftung;
the Department of Atomic Energy (Project Identification No. RTI 4002), the Department of Science and Technology of India,
and the UPES (India) SEED finding programs Nos. UPES/R\&D-SEED-INFRA/17052023/01 and UPES/R\&D-SOE/20062022/06; 
the Istituto Nazionale di Fisica Nucleare of Italy; 
National Research Foundation (NRF) of Korea Grant
Nos.~2016R1\-D1A1B\-02012900, 2018R1\-A2B\-3003643,
2018R1\-A6A1A\-06024970, RS\-2022\-00197659,
2019R1\-I1A3A\-01058933, 2021R1\-A6A1A\-03043957,
2021R1\-F1A\-1060423, 2021R1\-F1A\-1064008, 2022R1\-A2C\-1003993;
Radiation Science Research Institute, Foreign Large-size Research Facility Application Supporting project, the Global Science Experimental Data Hub Center of the Korea Institute of Science and Technology Information and KREONET/GLORIAD;
the Polish Ministry of Science and Higher Education and 
the National Science Center;
the Ministry of Science and Higher Education of the Russian Federation
and the HSE University Basic Research Program, Moscow; 
University of Tabuk research grants
S-1440-0321, S-0256-1438, and S-0280-1439 (Saudi Arabia);
the Slovenian Research Agency Grant Nos. J1-9124 and P1-0135;
Ikerbasque, Basque Foundation for Science, and the State Agency for Research
of the Spanish Ministry of Science and Innovation through Grant No. PID2022-136510NB-C33 (Spain);
the Swiss National Science Foundation; 
the Ministry of Education and the National Science and Technology Council of Taiwan;
and the United States Department of Energy and the National Science Foundation.
These acknowledgements are not to be interpreted as an endorsement of any
statement made by any of our institutes, funding agencies, governments, or
their representatives.
We thank the KEKB group for the excellent operation of the
accelerator; the KEK cryogenics group for the efficient
operation of the solenoid; and the KEK computer group and the Pacific Northwest National
Laboratory (PNNL) Environmental Molecular Sciences Laboratory (EMSL)
computing group for strong computing support; and the National
Institute of Informatics, and Science Information NETwork 6 (SINET6) for
valuable network support.

\providecommand{\href}[2]{#2}\begingroup\raggedright\endgroup

\newpage
\begin{appendices}

\section{Channels used to reconstruct $B$ and $D$ mesons in the FEI}
\label{appendixA}

\begin{table}[htbp]
 \caption{ Decay channels of $B_s^0$, $B^+$ and $B^0$ mesons used in the FEI. }
 \label{tab3}
\centering
\begin{tabular}{@{}lll@{}} \toprule
  $B_s^0\to$ & $B^+\to$ & $B^0\to$ \\
  \midrule
  $D_s^- \pi^+$ & $\bar{D}^0\pi^+$ & $D^-\pi^+$ \\
  $D_s^- \pi^+ \pi^0$ & $\bar{D}^0\pi^+\pi^0$ & $D^-\pi^+\pi^0$ \\
  $D_s^- \pi^+ \pi^+ \pi^-$ & $\bar{D}^0\pi^+\pi^+\pi^-$ & $D^-\pi^+\pi^+\pi^-$ \\
  $D_s^{*-} \pi^+$ & $\bar{D}^{*0}\pi^+$ & $D^{*-}\pi^+$ \\
  $D_s^{*-} \pi^0 \pi^+$ & $\bar{D}^{*0}\pi^+\pi^0$ & $D^{*-}\pi^+\pi^0$ \\
  $D_s^{*-} \pi^+ \pi^+ \pi^-$ & $\bar{D}^{*0}\pi^+\pi^+\pi^-$ & $D^{*-}\pi^+\pi^+\pi^-$ \\ 
  \midrule
   $D_s^- D_s^+$ & $D_s^{+}\bar{D}^{0}$ & $D_s^{+}D^{-}$ \\
   $D_s^{*-} D_s^+$ & $D_s^{*+}\bar{D}^{0}$ & $D_s^{*+}D^{-}$ \\
   $D_s^- D_s^{*+}$ & $D_s^{+}\bar{D}^{*0}$ & $D_s^{+}D^{*-}$ \\
  $D_s^{*-} D_s^{*+}$ & $D_s^{*+}\bar{D}^{*0}$ & $D_s^{*+}D^{*-}$ \\
  \midrule
  $J/\psi \, K^+ K^-$ & $J/\psi \, K^+$ & $J/\psi \, K_S^0$ \\
  $J/\psi \, K^+ K^- \pi^0$ & $J/\psi \, K_S^0\,\pi^+$ & $J/\psi \, K^+\pi^-$ \\
  & $J/\psi \, K^+\pi^+\pi^-$ & \\
  \midrule
  $\bar{D}^0 K^- \pi^+$ & $D^-\pi^+\pi^+$ & $D^{*-}K^+K^-\pi^+$ \\
  $\bar{D}^{*0} K^- \pi^+$ & $D^{*-}\pi^+\pi^+$ & \\
  $D_s^- K^+$ & & \\
  \bottomrule
 \end{tabular}
\end{table}

\begin{table}[htbp]
 \caption{ Decay channels of $D^0$, $D^+$ and $D_s^+$ mesons used in the FEI. }
 \label{tab4}
\centering
\begin{tabular}{@{}lll@{}} \toprule
  $D^0\to$ & $D^+\to$ & $D_s^+\to$ \\ \midrule
  $K^-\pi^+$ & $K^-\pi^+\pi^+$ & $K^+K^-\pi^+$ \\
  $K^-\pi^+\pi^0$ & $K^-\pi^+\pi^+\pi^0$ & $K^+K_S^0$ \\
  $K^-\pi^+\pi^+\pi^-$ & $K_S^0\,\pi^+$ & $K^+K^-\pi^+\pi^0$ \\
  $K_S^0\,\pi^+\pi^-$ & $K_S^0\,\pi^+\pi^0$ & $K^+K_S^0\,\pi^+\pi^-$ \\
  $K_S^0\,\pi^+\pi^-\pi^0$ & $K_S^0\,\pi^+\pi^+\pi^-$ & $K^-K_S^0\,\pi^+\pi^+$ \\
  $K^+K^-$ & $K^+K^-\pi^+$ & $K^+K^-\pi^+\pi^+\pi^-$ \\
  $K^+K^-K_S^0\,$ & & $K^+\pi^+\pi^-$ \\
  & & $\pi^+\pi^+\pi^-$ \\
   & & $ K^+ K_S^0 \pi^0$\\
  & & $K_S^0 K_S^0 \pi^+$\\
   & & $\eta \pi^+$\\
   & & $\eta' \pi^+$\\
   & & $\eta \pi^+ \pi^0$\\
   & & $\eta' \pi^+ \pi^0$\\
   
  \bottomrule
 \end{tabular} 
\end{table}

\section{Determination of weights for simulated events}
\label{Weights}
We introduce weights for simulated events to take into account the difference between relative yields of $B_s^0$ channels in simulation and data. To determine the weights, we use the $\mbs$ distributions in the $\Upsilon(5S)$ sample. We select $B_s^0$ candidates using requirements on their momenta measured in the center-of-mass frame, $p^*(B_s)$, and the signal probability, ${\cal P}_{B_s}$, which are close to the optimal requirements for measuring $\Br(B_s^0 \to D_s^{\pm}X)$. We perform a simultaneous fit to the $\mbs$ distributions in data and
simulation for each $B_s^0$ decay channel. The signal in simulation is described by a sum of three Gaussians; the signal in data is described by the same Gaussians with additional free parameters representing overall normalization, a shift in the means, and a broadening of the Gaussian widths to adjust for differences between data and simulation. The background in simulation and in data is described by a second order polynomial. The weights are typically in the range $0.5-1.5$.

\section{{Broken signal and peaking background calibration}}
\label{appendixPBG}
We calibrate the simulation of the broken-signal components due to the loss of a photon from the $D_s^* \to D_s \gamma$ decay and misidentification of a kaon as a pion, as well as the $B^0$ peaking background. All these peaking structures are situated in the left $M(B_s)$ sideband.

\subsection{Loss of a photon from $D_s^*$}
\begin{figure}[htbp]
	\centering
	\includegraphics[width = 0.7\textwidth]{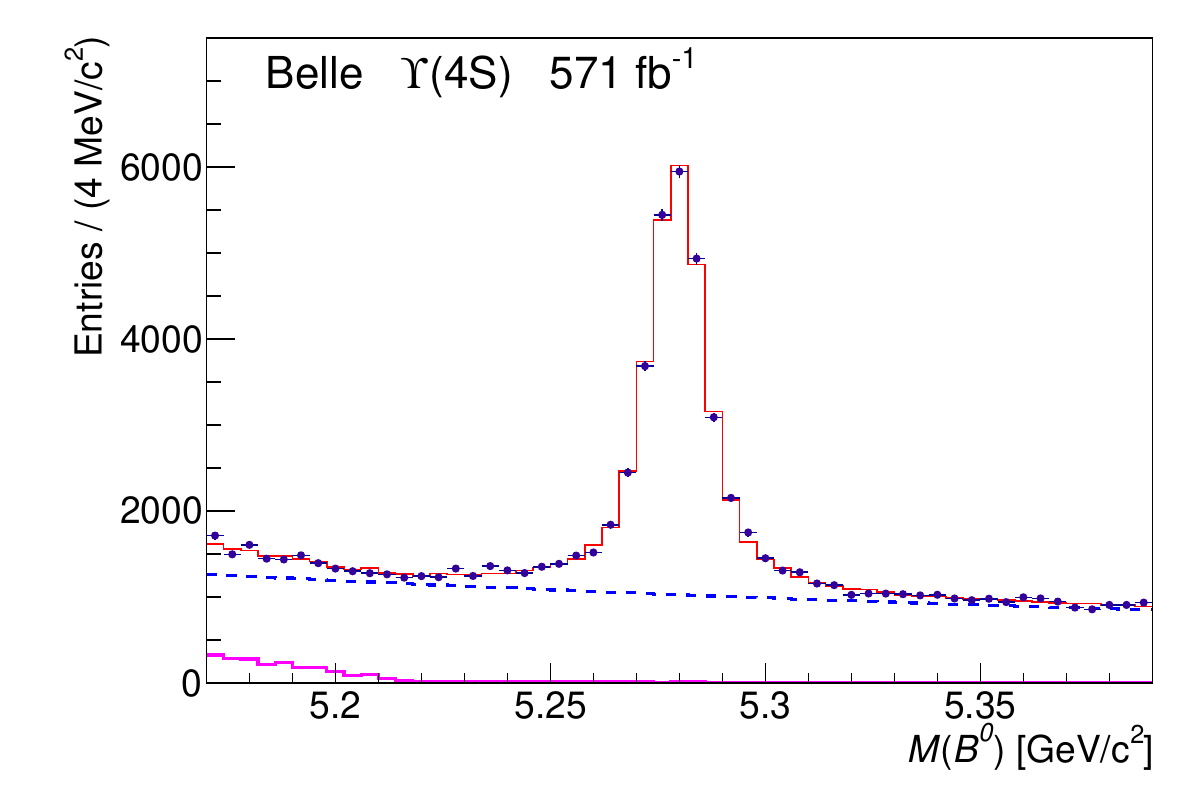}
	\caption{The mass distribution for the selected $B^0 \to D_s^+ D^-$ candidates. The points with error bars are data, the solid red histogram is the result of the fit, the solid magenta histogram is the broken-signal component caused by the loss of a photon, and the dashed blue histogram is the smooth background component of the fit function.\label{figGamma}}
\end{figure}

The ratio of the number of broken-signal events due to misreconstructed $D_s^* \to D_s \gamma$ to the number of events in the $B_s^0$ peak is determined from simulation to be 9--11\%. This ratio is multiplied by the calibration factor to adjust for differences between data and simulation in these effects. This calibration factor is obtained using the $M(B^0)$ distribution of $B^0$ candidates reconstructed in part of the $\Upsilon(4S)$ data with an integrated luminosity of 571 fb$^{-1}$. The $B^0$ decay channels used in the FEI are listed in table~\ref{tab3} in Appendix~\ref{appendixA}. We use the mass distribution of the candidates reconstructed in $B^0 \to D_s^+ D^-$ channel. Introducing the peaking background component and fitting to the mass distribution in data, we find a calibration factor of $0.75 \pm 0.25$. The result of this fit is shown in figure~\ref{figGamma}. The calibrated relative normalization of this broken-signal component, which depends on the ${\cal P}_{B_s}$ requirement, is shown in table~\ref{PBG_norm}.

 \begin{table}[htbp]
 	\caption{{The broken-signal (BS) yields relative to the $B_s^0$ signal yield and the yields of $B^0$ peaking background for various requirements on the $B_s$ signal probability ${\cal P}_{B_s}$.}}
		\label{PBG_norm}
		\begin{center}
    \begin{tabular}{@{}lcccc@{}}

  \toprule

  Decay & ${\cal P}_{B_s}$ cut & $D_s^*$ BS, \% & misID BS, \% & Number of $B^0$ \\
  \midrule
  $B_s^0 \to D_s^{\pm} X$ & $> 0.0012$ & $8.4 \pm 2.8$ & $0.89 \pm 0.02$ & $131 \pm 9$ \\
   $B_s^0 \to D^0/\bar{D}^0 X$ & $> 0.0050$ & $7.5 \pm 2.5$ & $0.85 \pm 0.02$ & $97 \pm 7$\\
	 $B_s^0 \to D^{\pm} X$ & $> 0.0200$ & $6.7 \pm 2.2$ & $0.79 \pm 0.01$ & $57 \pm 5$\\
  \bottomrule
   \end{tabular}
		\end{center}

	\end{table}

\subsection{Misidentification of a kaon}
The normalization of the misidentification component is 0.75--0.85\% of the number of events in the $B_s^0$ peak. The difference between MC and data in kaon misidentification rate is studied using $D^{*+} \to D^0(K^- \pi^+)\pi^+$ decays~\cite{BaBar:2014omp}. The calibration factor for the broken signal due to kaon misidentification is found to be $1.06\pm0.02$. The calibrated relative normalization of this broken signal for various ${\cal P}_{B_s}$ requirements is shown in table~\ref{PBG_norm}.

\subsection{Reconstruction of a $B^0$ meson as $B_s^0$}
\begin{figure}[htbp]
	\centering
	\includegraphics[width = 0.7\textwidth]{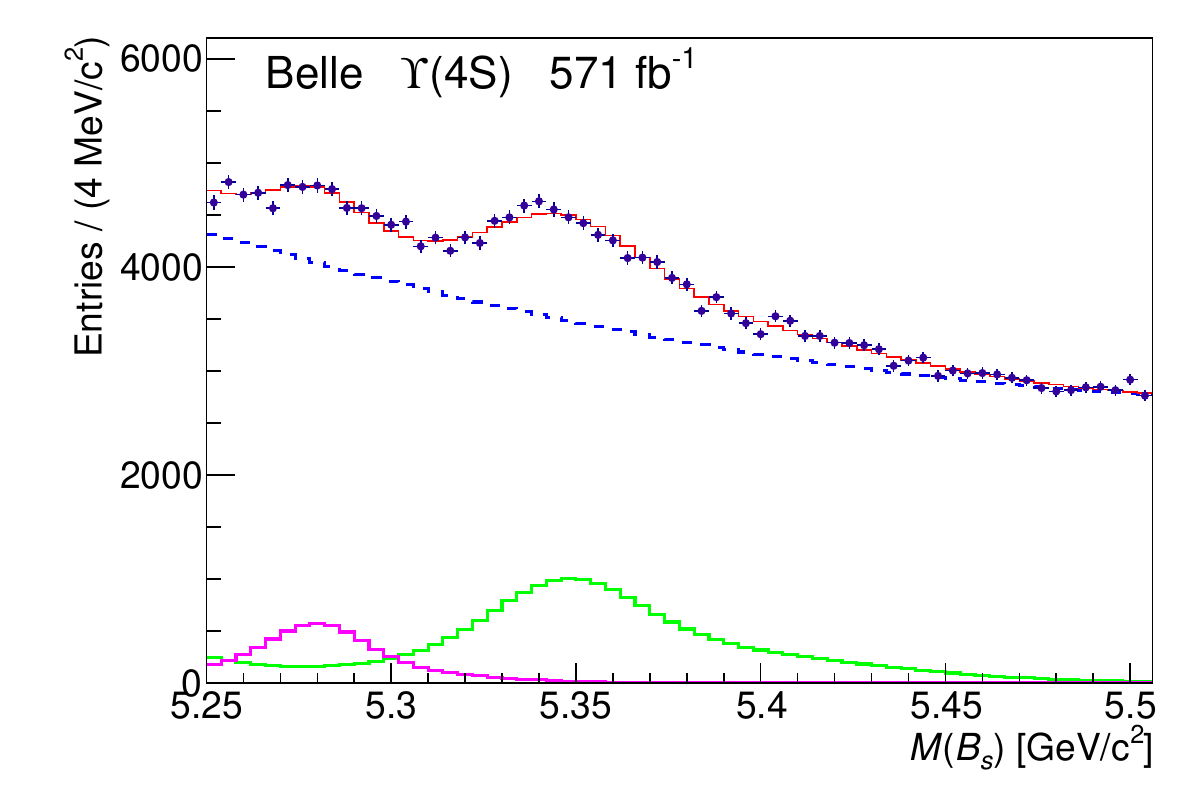}
	\caption{The mass distribution for the selected $B_s^0$ candidates at the $\Upsilon(4S)$ resonance. The points with error bars are data, the solid red histogram is the result of the fit, the solid green histogram is the component corresponding to correct reconstruction of all $B^0$ decay products, and the solid magenta histogram corresponds to the $B^0$ candidates in which daughter pion was misidentified as a kaon. The dashed blue histogram is the smooth background component of the fit function.\label{figB0}}
\end{figure}

  To satisfy our momentum selection criteria, $|p_{cm} - p^*(B_s)| < 0.09$, $B^0$ mesons must be produced in $e^+ e^- \to \Upsilon(4S)(\to B^0 \bar{B}^0)\, \gamma_{ISR}$ or $e^+ e^- \to B^{*} \bar{B}^* \pi$ processes whose cross-section in simulation may differ from that in data. To eliminate this uncertainty, we use part of the $\Upsilon(4S)$ data with an integrated luminosity of 571 fb$^{-1}$, in which events with the $B^0$ meson misreconstructed as $B_s^0$ also take place. The number of these events is obtained fitting to the $M(B_s)$ distribution and found to be $36670 \pm 2110$, $27150 \pm 1220$, and $16020 \pm 650$, depending on the ${\cal P}_{B_s}$ requirement. The result of the fit when the ${\cal P}_{B_s} > 0.005$ selection requirement is applied is shown in figure~\ref{figB0}. The number of signal $B^0$ reconstructed in the same data sample in channels shown in table~\ref{tab3} is equal to $439530 \pm 950$. At the $\Upsilon(5S)$ we reconstruct $1565 \pm 130$ signal $B^0$ satisfying the $p^*(B)$ selection criteria. Assuming that the $B^0$ reconstruction efficiency does not depend on the $B^0$ momentum, we obtain the expected number of fake $B_s^0$ candidates at the $\Upsilon(5S)$ resonance given in table~\ref{PBG_norm}.

\end{appendices}

\end{document}